\documentclass[conference]{IEEEtran}
\pdfoutput=1
\IEEEoverridecommandlockouts
\usepackage{cite}
\usepackage{amsmath,amssymb,amsfonts}
\usepackage{graphicx}
\usepackage{textcomp}
\usepackage{xcolor}
\usepackage{subfig}
\usepackage[T1]{fontenc}

\begin{document}

\title{Fast ultrasonic imaging using end-to-end deep learning\\
\thanks{This work was supported by Applus+ RTD, CWI and the Dutch Research Council (NWO 613.009.106, 639.073.506). Submission - IEEE International Ultrasonics Symposium 2020.} }
\author{
    \IEEEauthorblockN{Georgios Pilikos\IEEEauthorrefmark{1}, Lars Horchens\IEEEauthorrefmark{2}, Kees Joost Batenburg\IEEEauthorrefmark{1}\IEEEauthorrefmark{3}, Tristan van Leeuwen\IEEEauthorrefmark{1}\IEEEauthorrefmark{10} and Felix Lucka\IEEEauthorrefmark{1}\IEEEauthorrefmark{4}}
    \IEEEauthorblockA{\IEEEauthorrefmark{1}Computational Imaging, Centrum Wiskunde \& Informatica, Amsterdam, NL
    }
    \IEEEauthorblockA{\IEEEauthorrefmark{2}Applus+ E\&I Technology Centre, Rotterdam, NL}
    \IEEEauthorblockA{\IEEEauthorrefmark{3}Leiden Institute of Advanced Computer Science, Leiden University, Leiden, NL}
    \IEEEauthorblockA{\IEEEauthorrefmark{10}Mathematical Institute, Utrecht University, Utrecht, NL}
    \IEEEauthorblockA{\IEEEauthorrefmark{4}Centre for Medical Image Computing, University College London, London, UK}
}

\maketitle

\begin{abstract}
Ultrasonic imaging algorithms used in many clinical and industrial applications consist of three steps: A data pre-processing, an image formation and an image
post-processing step. For efficiency, image formation often relies on an approximation of the underlying
wave physics. A prominent example is the Delay-And-Sum (DAS) algorithm used in reflectivity-based ultrasonic imaging. Recently, deep neural networks (DNNs) are being used for the data pre-processing and the image post-processing steps separately. In this work, we propose a novel deep learning architecture that integrates all three steps to enable end-to-end training. We examine turning the DAS image formation method into a network layer that connects data pre-processing layers with image post-processing layers that perform segmentation. We demonstrate that this integrated approach clearly outperforms sequential approaches that are trained separately. While network training and evaluation is performed only on simulated data, we also showcase the potential of our approach on real data from a non-destructive testing scenario.
\end{abstract}


\section{Introduction}
Ultrasonic imaging aims at generating maps of the acoustic properties of a medium of interest. It has certain advantages over other imaging modalities such as magnetic resonance imaging (MRI) or computed tomography (CT): it uses non-ionizing radiation, it is mobile, has low operating costs and enables real-time imaging \cite{ultrafast}. Nevertheless, the compromise in achieving fast and interactive imaging is that the resulting images require substantial human expertise for their interpretation and differentiating between materials is not trivial.

Typical workflows for 2D ultrasonic imaging with linear arrays consist of three steps: (1) data pre-processing (e.g. denoising, filtering, deconvolution), (2) image formation via beamforming and (3) image post-processing (e.g. image enhancement or segmentation). However, this three-step process introduces data/reconstruction errors which propagate due to inaccurate physics modelling or noise in the data.

Recently, there have been efforts to implement these steps with deep learning \cite{dlus}. Deep neural networks (DNNs) use raw data as input and output an image \cite{acb} - \cite{simson}. The final goal is not usually to produce an image but rather it is an intermediate step before image enhancement or segmentation. Further work utilizes two decoders to obtain a beamformed image and a segmentation from one encoder using raw data \cite{segmNair}. Nonetheless, integrating the image formation, which approximates the underlying wave physics, within the deep learning architecture has shown to improve final results\cite{ongie2020deep} \cite{arridge}.

In this work, we propose to integrate all three steps together to enable end-to-end training. To achieve this, we propose a novel architecture that utilizes a fast ultrasonic imaging operator, the Delay-And-Sum (DAS). We examine turning the DAS image formation into a network layer that connects data pre-processing and image post-processing DNNs. Using this, we propose an end-to-end training strategy to obtain improved results. In section 2, we describe the ultrasonic data acquisition and the DAS image formation. Then, in section 3, we introduce our proposed end-to-end deep learning approach and in section 4, we demonstrate that our integrated approach outperforms sequential approaches which are trained independently using simulated data. Finally, we apply this to real data showcasing its potential to a non-destructive testing scenario. 
\begin{figure}
\centering
\includegraphics[scale=0.2]{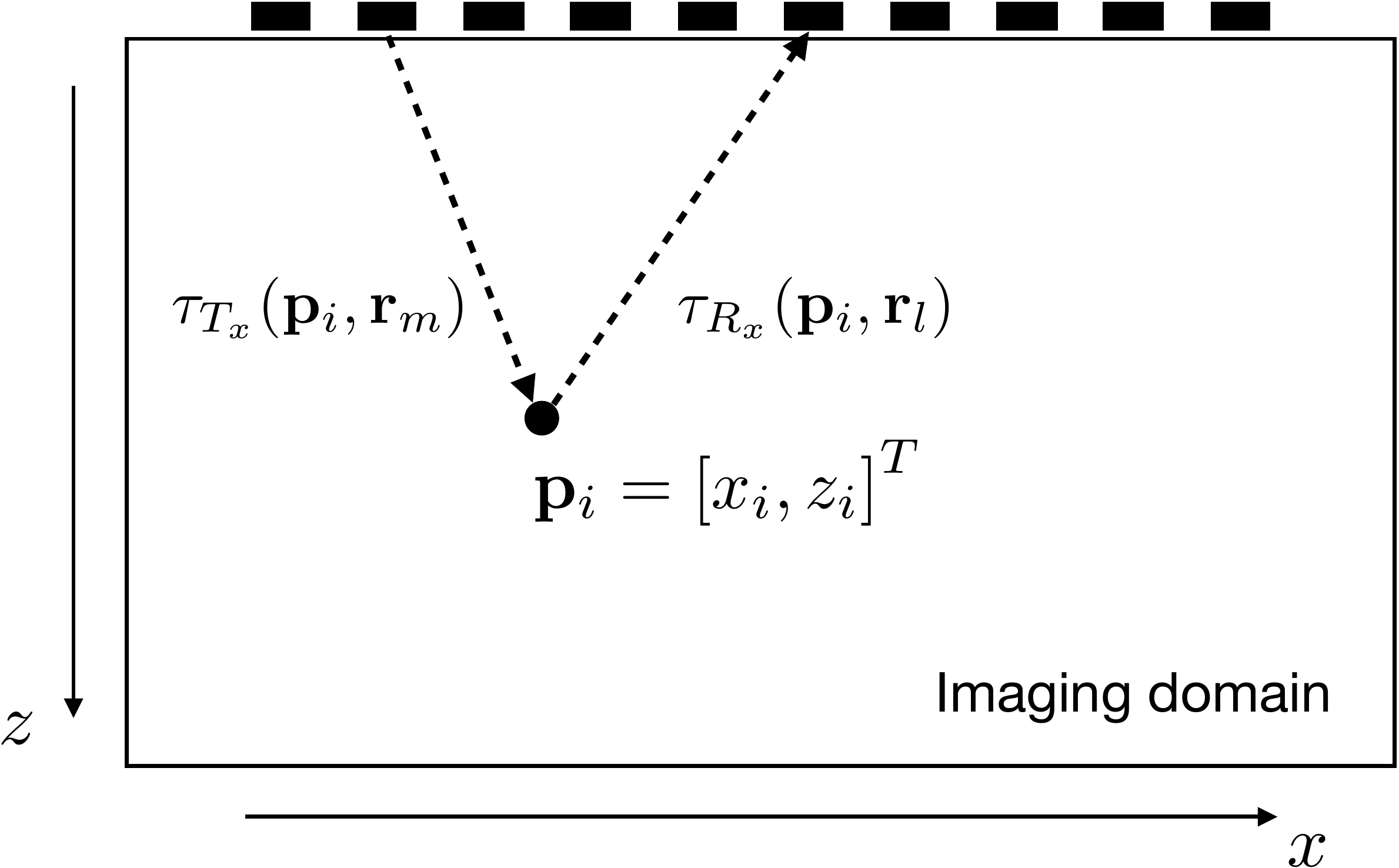}
\caption{A phased-array with source/receive elements ($\mathbf r$) depicted as black rectangles. For each image point, $\mathbf p_i$, and each source/receiver combination, travel times are calculated.}
\label{setup}
\end{figure}

\begin{figure*}
\centering
\includegraphics[scale=0.124656]{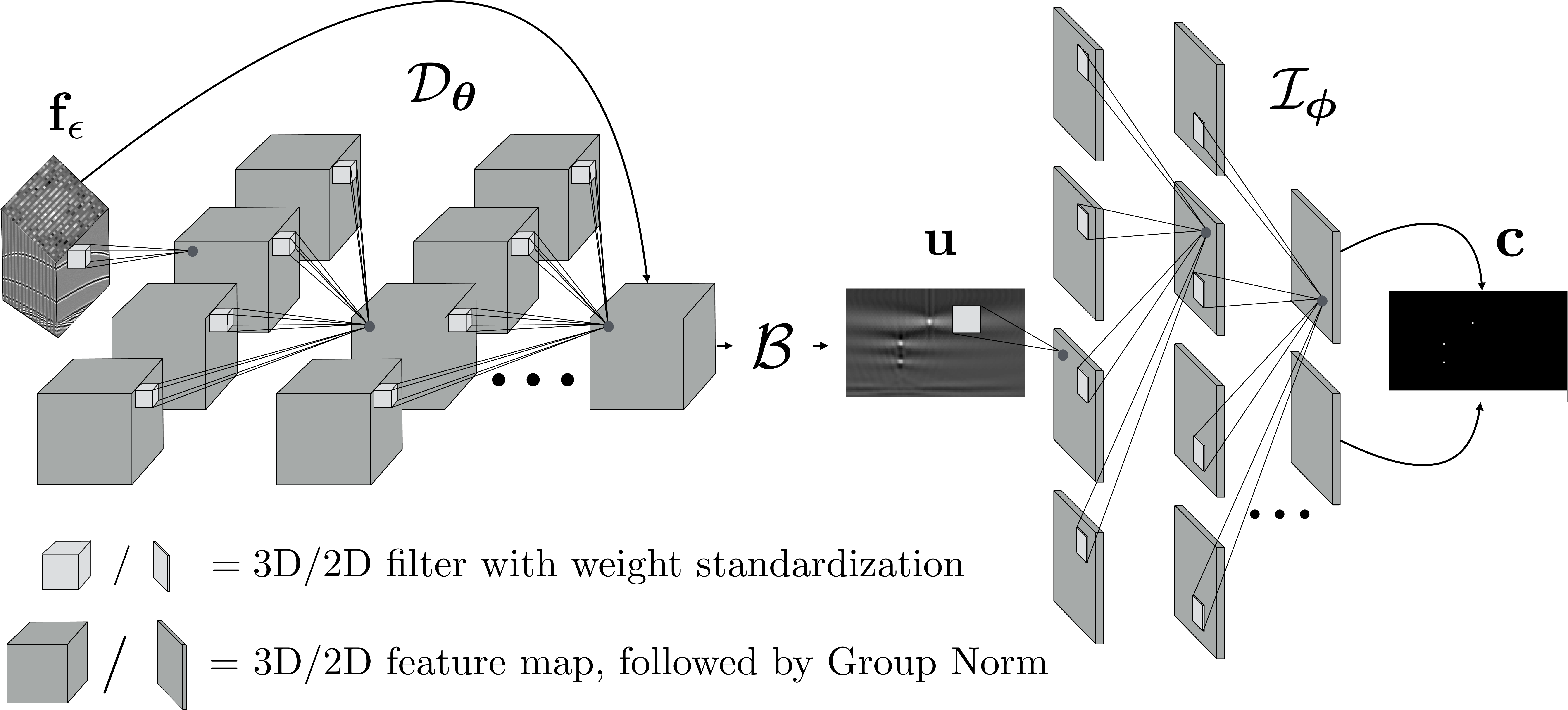}
\caption{A 3D DCNN, $\mathcal{D}_{\boldsymbol \theta}$, is used for data pre-processing and the DAS operator, $\mathcal B$, incorporates the image formation into the whole network. A 2D DCNN, $\mathcal{I}_{\boldsymbol \phi}$, post-processes the intermediate image and produces the final result. Feature maps are followed by Group Norm and ReLU (tanh is used at the last layer of $\mathcal{D}_{\boldsymbol \theta}$). Only one filter at one location per layer is shown.}
\label{e2e}
\end{figure*}

\section{Ultrasonic imaging}
We examine 2D data acquisition with a linear array as shown in Figure \ref{setup}. An element is used as a source, $\mathbf r_m$, and transmits a pulsed ultrasonic wave into the medium of interest. All receivers capture the resulting wave field, which contains information about the wave-matter interactions, e.g. via reflections from interfaces with different acoustic properties. The data acquisition continues with the next element as a source and so on until all elements have acted as sources \cite{tfm},\cite{iwex}. 

This is called Full Matrix Capture (FMC) and leads to a data volume, $ \mathbf f \in \mathbb{R}^{n_t \times n_s \times n_r}$, where $n_t$, $n_s$ and $n_r$ are the number of time samples, sources and receivers respectively. The aim is to obtain an image, $\mathbf u \in \mathbb{R}^{n_x \times n_z}$, where $n_x$ and $n_z$ are the number of pixels in the horizontal and vertical direction. 

\subsection*{Delay-And-Sum image formation}
Delay-And-Sum (DAS) relies on an approximation of the underlying wave physics. It calculates travel times, $\tau(\mathbf p_i, \mathbf r_m, \mathbf r_l)$, between each source, $\mathbf r_m$, each image point, $\mathbf p_i = \left[x_i, z_i\right]^{T}$ and each receiver, $\mathbf r_l$, assuming a uniform speed of sound in the material, $s$. This is calculated by,
\begin{equation}
\tau(\mathbf p_i, \mathbf r_m, \mathbf r_l) = \frac{\|\mathbf r_m - \mathbf p_i \|_2}{s} +  \frac{\|\mathbf r_l - \mathbf p_i \|_2}{s},
\end{equation}
and depicted in Figure \ref{setup}. Each travel time is converted into an index using the sampling frequency which is used to locate a sample. This time-shift operation corresponds to the \emph{delay} part. The amplitude is extracted at that time-shifted location, and the process is repeated for all travel times corresponding to an image point. Finally, it \emph{sums} all amplitudes giving image amplitude, $u_i$, for image point, $\mathbf p_i$. This can be written as,
\begin{equation}
u_i = \sum_{m=0}^{n_s} \sum_ {l=0}^{n_r}f(\tau(\mathbf p_i, \mathbf r_{m}, \mathbf r_{l}),m,l) ,
\end{equation} and repeated for all image points to form an image. We can write it as a linear operator, $\mathcal{B}: \mathbb{R}^{n_t \times n_s \times n_r}  \to  \mathbb{R}^{n_x 
\times n_z} $, and referred as \emph{DAS operator}. The whole process is written as,
\begin{equation}\label{das}
\mathbf u = \mathcal{B}\mathbf{f}.
\end{equation}  

\begin{figure*}
\centering
\subfloat[]{ 
	\includegraphics[scale=0.38]{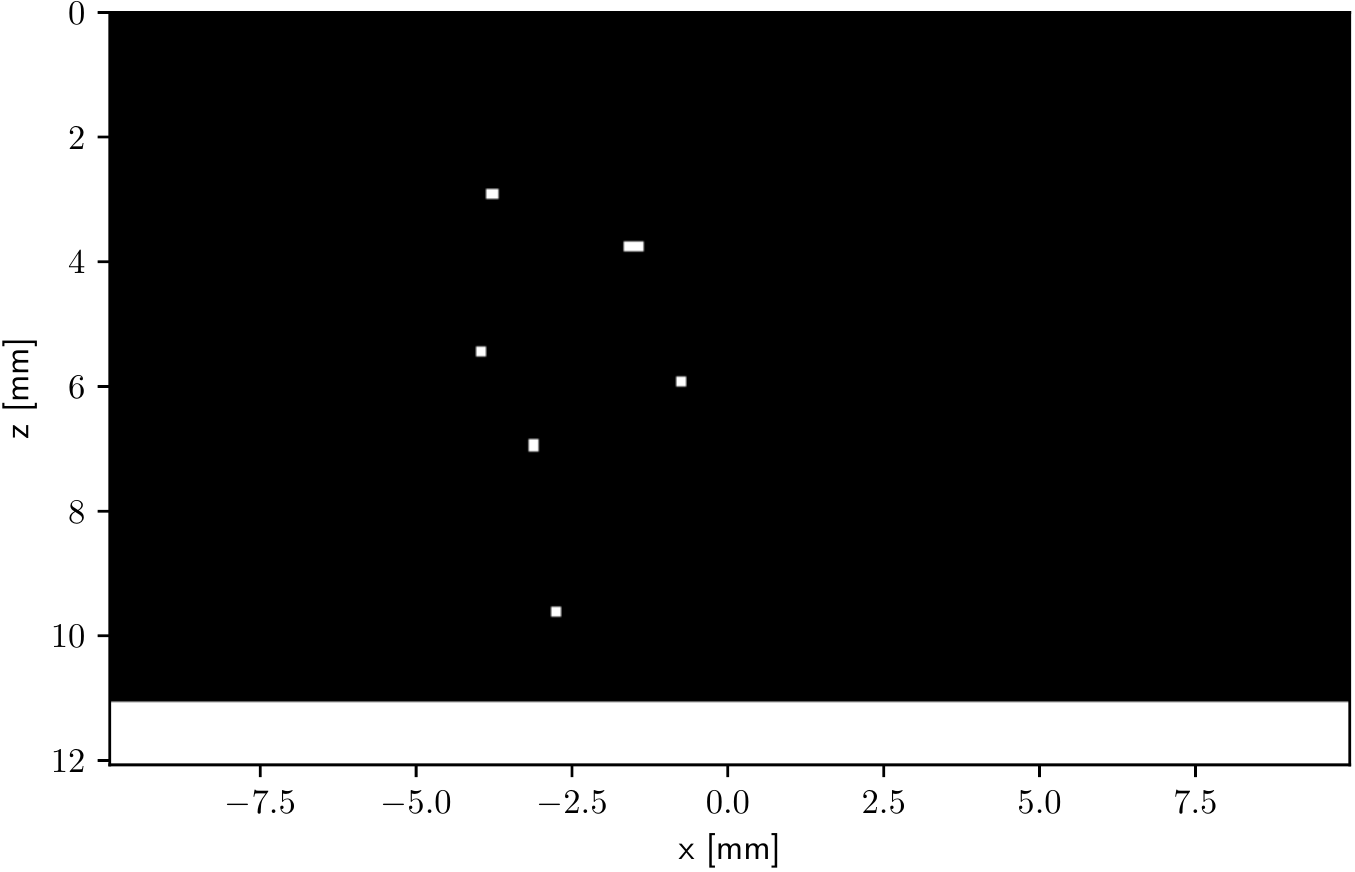}
	}
\subfloat[]{ 
	\includegraphics[scale=0.38]{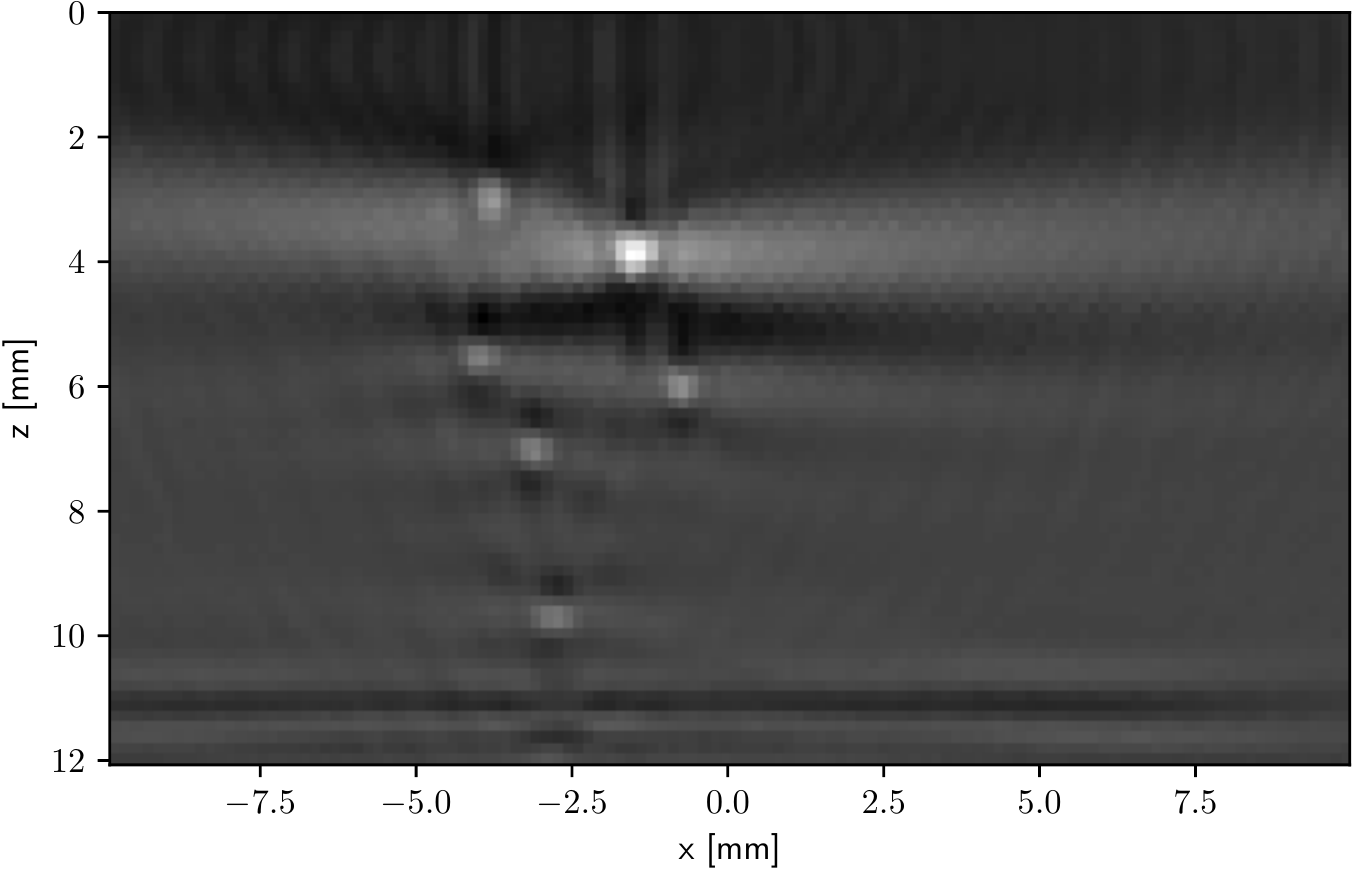}
	}
\subfloat[]{
	\includegraphics[scale=0.38]{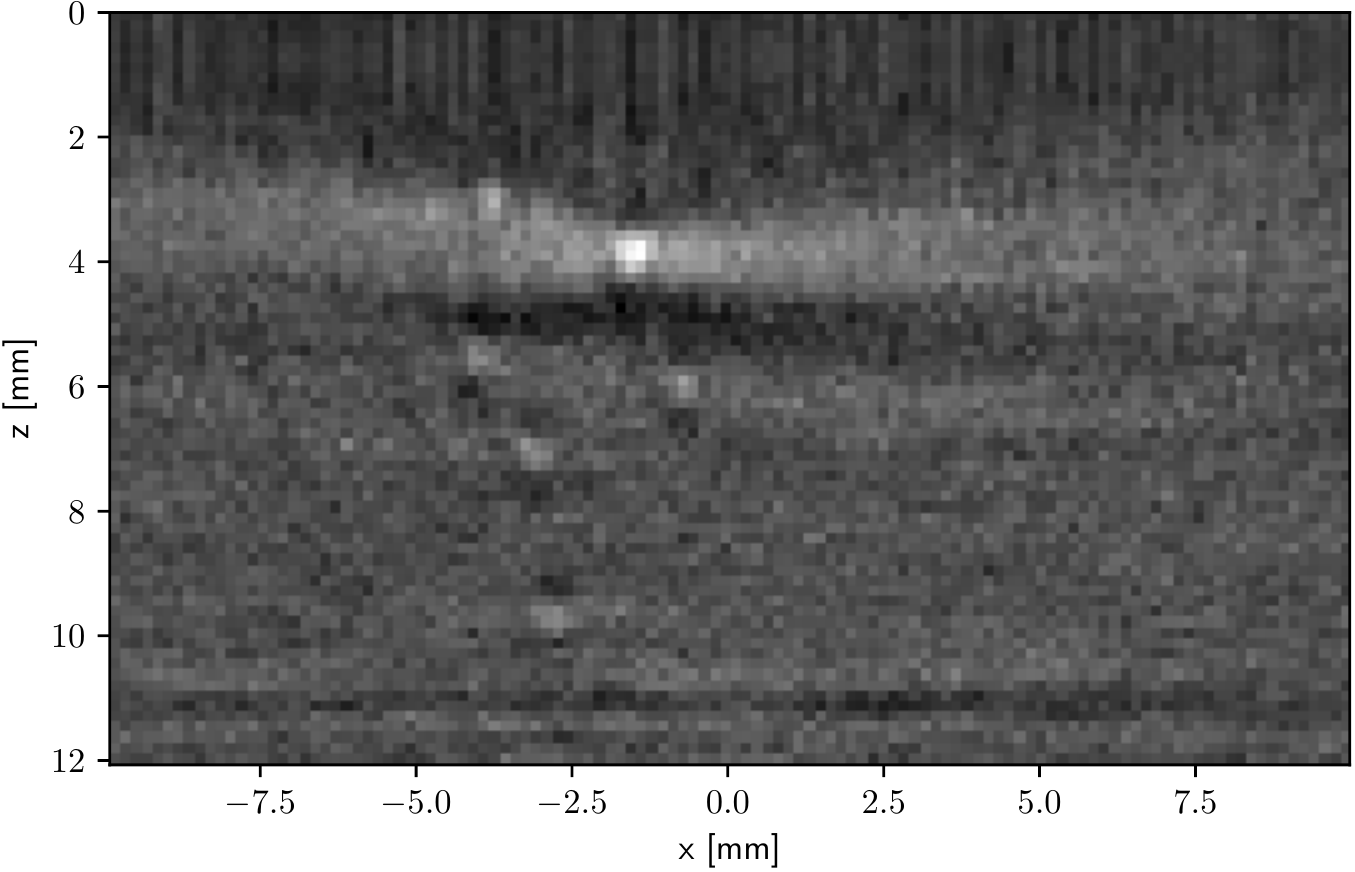}
	}	\\
\subfloat[]{ 
	\includegraphics[scale=0.38]{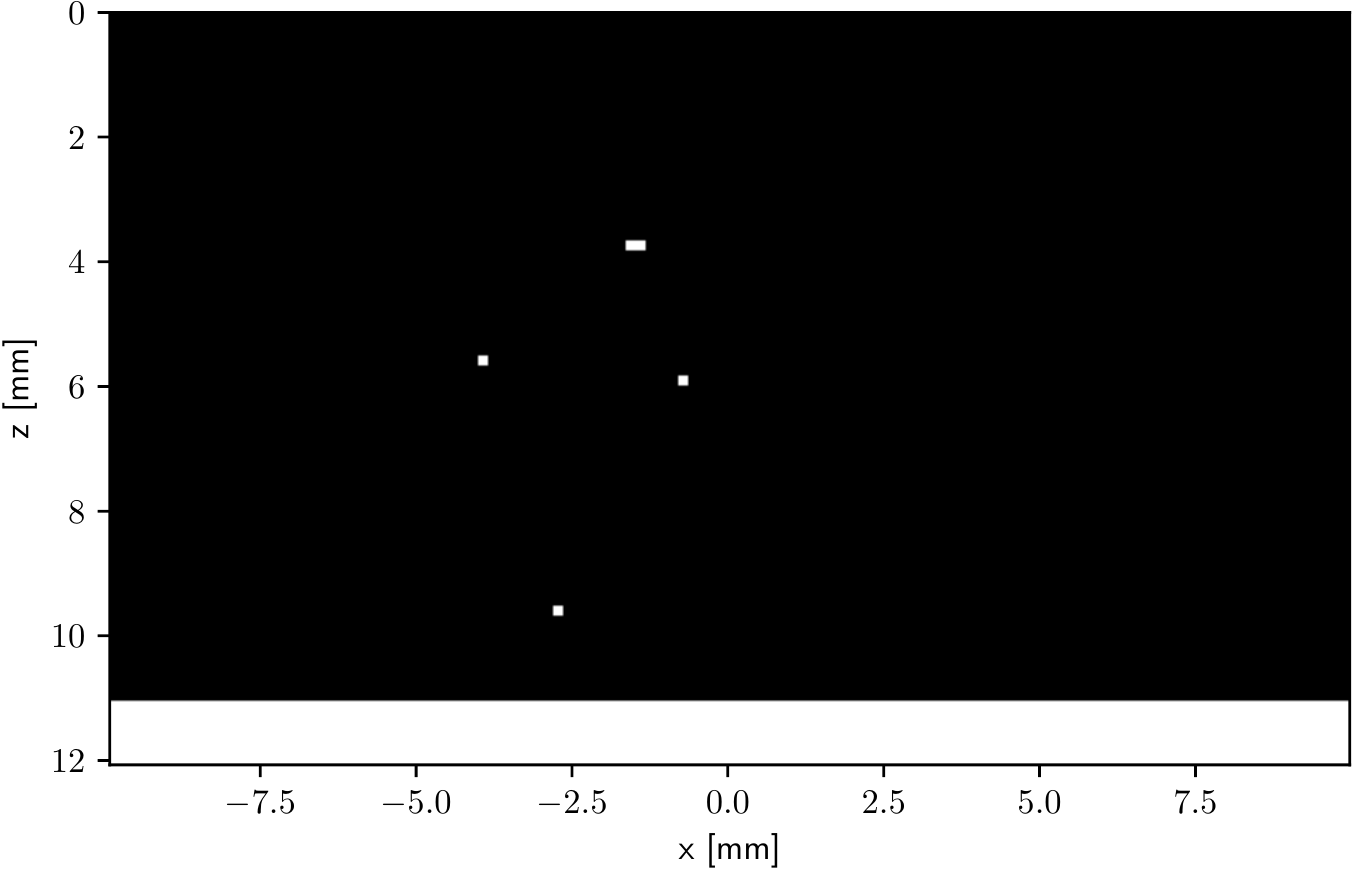}
	}
\subfloat[]{ 
	\includegraphics[scale=0.38]{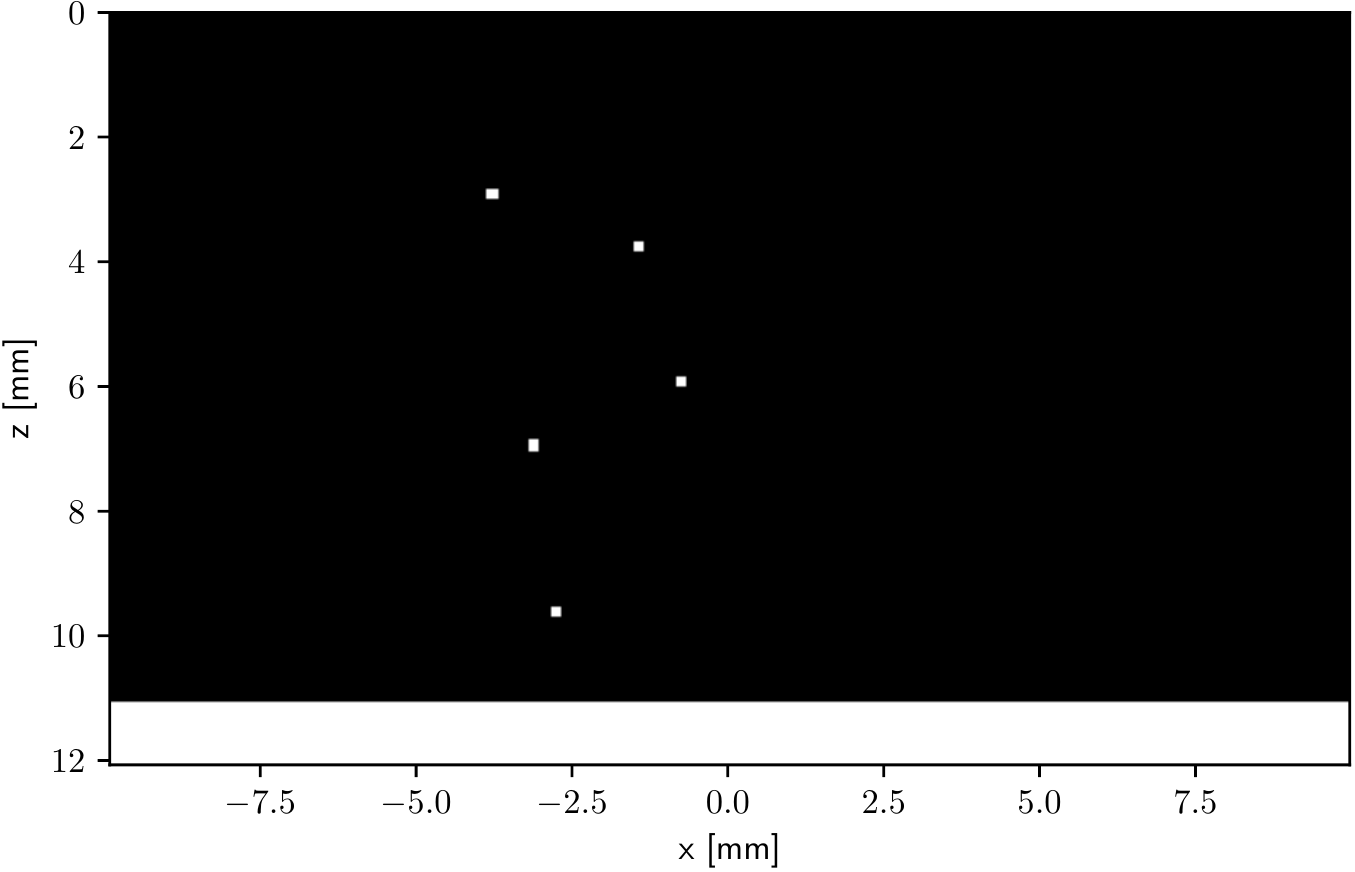}
	}
\subfloat[]{
	\includegraphics[scale=0.38]{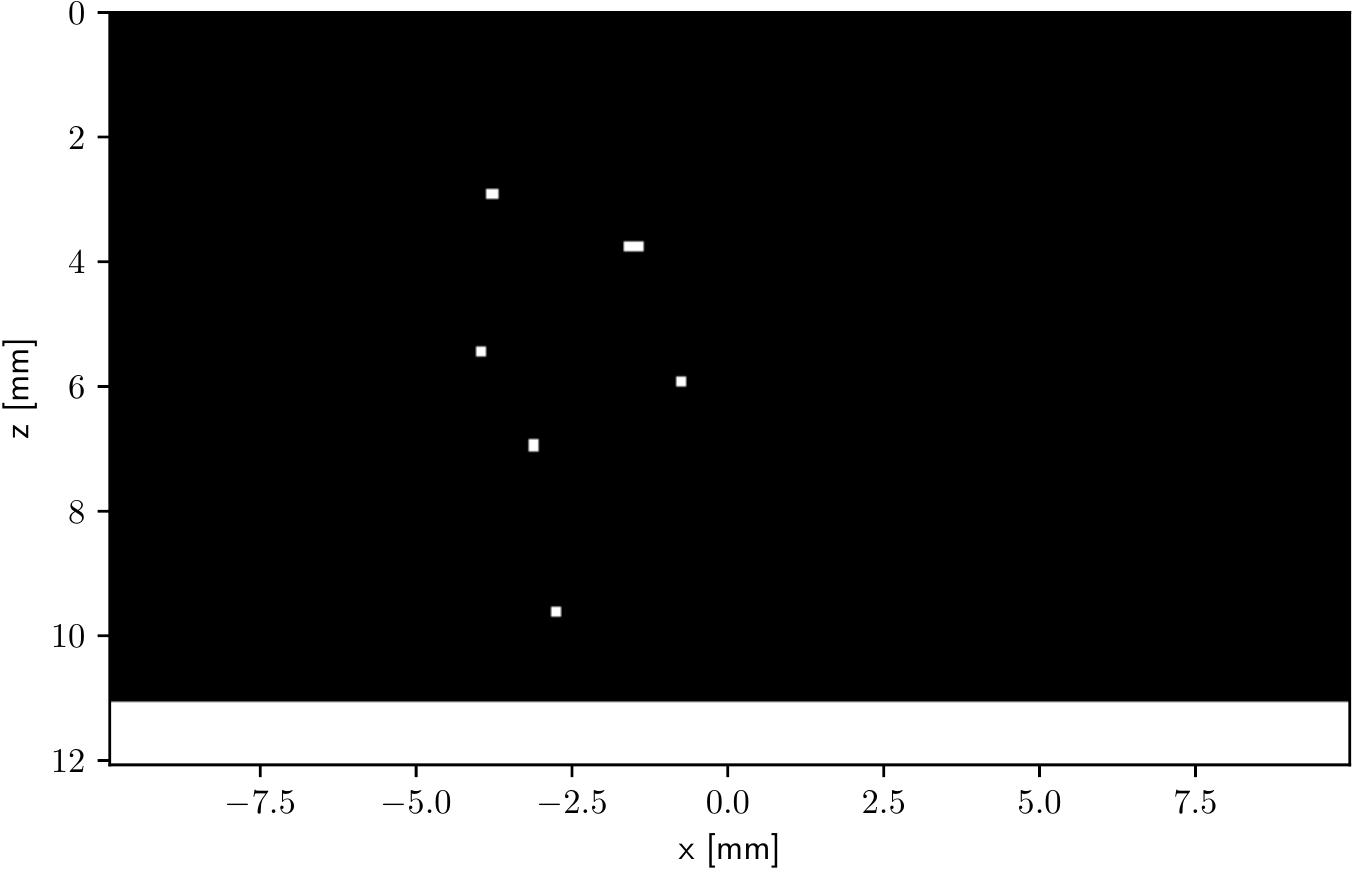}
	}	
\caption{(a) Target segmentation (white: air, black: carbon steel), (b) DAS image from fully-sampled data, (c) DAS image from noisy and under-sampled data, (d)-(f) segmentation result of 1st/2nd/3rd training strategy from noisy and under-sampled data.}
\label{simulated}
\end{figure*}

Note that the true mapping from acoustic properties to data is non-linear and includes a lot of different, complicated wave-matter interactions. On the other hand, the DAS algorithm corresponds to a linear back-projection-type operator that tries to form an approximate, qualitative image of acoustic property variations in space. Due to this approximation, it is usually preceded by data pre-processing and followed by image post-processing. These individual steps are increasingly being replaced by deep convolutional neural networks (DCNNs) \cite{dlus}.

\section{End-to-end deep learning}
DCNNs are parameterized non-linear mappings optimized for a given loss function. In this work, we propose a novel architecture, as shown in Figure \ref{e2e}. The architecture involves a 3D DCNN, which is a data-to-data mapping $\mathcal{D}_{\boldsymbol \theta}$, acting on the data volume. Then, we incorporate the DAS operator, $\mathcal{B}$, into the network, by implementing a layer that applies the DAS algorithm on the data. For this, we also need to allow backpropagation of errors during training by deriving and implementing its adjoint action. Finally, the intermediate image formed is processed by a 2D DCNN, which is an image-to-image mapping $\mathcal{I}_{\boldsymbol \phi}$, to obtain a segmented image. This enables end-to-end training of DCNN parameters $\boldsymbol \theta$ and $\boldsymbol \phi$ simultaneously. Both DCNNs have 4 layers with 4 filters, each with $5\times5\times5$ and $5\times5$ dimensions. Weight Standardization \cite{ws} and Group Normalization \cite{gn} is used per layer to help training stability since we use one training sample per mini-batch. Furthermore, skip connections in $\mathcal{D}_{\boldsymbol \theta}$ enable better information flow and reduce training time \cite{unet}.

\subsection{Training strategies}
We will examine two sequential training strategies and introduce our proposed end-to-end approach. To facilitate discussion, we define $\mathbf c^{(i)}$ as the ground truth segmentation, $\mathbf f^{(i)}$ as clean simulated data, $\mathbf f^{(i)}_\epsilon$ as noisy, undersampled data and $\mathbf u^{(i)}$ as the DAS image from $\mathbf f^{(i)}$ using equation \ref{das}. The superscript $i$ represents the $i$-th training sample from a collection of training data, $\{\mathbf c^{(i)}, \mathbf f^{(i)}, \mathbf f^{(i)}_\epsilon,  \mathbf u^{(i)} \}_{i=1}^N$. All training strategies use the same loss function for image post-processing, namely the cross entropy loss referred to as $\mathcal H$ hereafter. The strategies are:

\emph{1st training strategy:} 
Data pre-processing DCNN, $\mathcal D_{\boldsymbol \theta}$, is trained and fixed. The DAS operator, $\mathcal B$, is applied to pre-processed data to form images. Then, image post-processing DCNN, $\mathcal I_{\boldsymbol \phi}$, is trained using these images. That is,
\begin{enumerate}
\item train $\boldsymbol{\skew{2}\hat\theta} := \underset{\boldsymbol \theta}{\operatorname{argmin}} \sum_{i=1}^N \| \mathbf f^{(i)} - \mathcal{D}_{\boldsymbol \theta}(\mathbf f^{(i)}_\epsilon) \|^2_2$
\item compute $\hat{\mathbf u}^{(i)} := \mathcal{B}\mathcal{D}_{\boldsymbol{\skew{2}\hat\theta}}(\mathbf f^{(i)}_\epsilon)$
\item train $\boldsymbol{\skew{3}\hat\phi} := \underset{\boldsymbol \phi}{\operatorname{argmin}} \sum_{i=1}^N \mathcal{H}(\mathbf c^{(i)}, \mathcal{I}_{\boldsymbol \phi}(\hat{\mathbf u}^{(i)}))$
\end{enumerate}

\emph{2nd training strategy:}
Data pre-processing DCNN, $\mathcal D_{\boldsymbol \theta}$, and DAS operator, $\mathcal B$, are trained together. Then, image post-processing DCNN, $\mathcal I_{\boldsymbol \phi}$, is trained. That is,
\begin{enumerate}
\item train $\boldsymbol{\skew{2}\hat\theta} := \underset{\boldsymbol \theta}{\operatorname{argmin}} \sum_{i=1}^N \| \mathbf u^{(i)} - \mathcal{B} \mathcal{D}_{\boldsymbol \theta}(\mathbf f^{(i)}_\epsilon) \|^2_2$
\item compute $\hat{\mathbf u}^{(i)} := \mathcal{B}\mathcal{D}_{\boldsymbol{\skew{2}\hat\theta}}(\mathbf f^{(i)}_\epsilon)$
\item train $\boldsymbol{\skew{3}\hat\phi} := \underset{\boldsymbol \phi}{\operatorname{argmin}} \sum_{i=1}^N \mathcal{H}(\mathbf c^{(i)}, \mathcal{I}_{\boldsymbol \phi}(\hat{\mathbf u}^{(i)}))$
\end{enumerate}

\emph{3rd training strategy:}
All three steps are combined and trained together in an end-to-end way as proposed. That is,
\begin{enumerate}
\item train $(\boldsymbol{\skew{3}\hat\phi}, \boldsymbol{\skew{2}\hat\theta}) := \underset{(\boldsymbol \phi, \boldsymbol \theta)}{\operatorname{argmin}} \sum_{i=1}^N \mathcal{H}(\mathbf c^{(i)}, \mathcal{I}_{\boldsymbol \phi}(\mathcal{B}\mathcal{D}_{\boldsymbol \theta}(\mathbf f^{(i)}_\epsilon)))$
\end{enumerate}

For the 2nd and 3rd strategies, we initialize the DCNNs with the parameters learnt by 1st and 2nd strategies respectively. 
  
\section{Experiments}
To evaluate our proposed approach, we use an ultrasonic-based non-destructive inspection of pipelines for defects.

\subsection{Simulated data}
The data domain was set to $64\times 64\times 1020$ with $64$ elements, $1020$ time samples and sampling frequency of $50\text{MHz}$. The image domain was set to $72\times 354$ pixels with defects randomly located around the middle of the domain. This was then cropped to $72\times 118$, as used in the real data acquired. As a proof of concept, we set the number of materials to 2. The segmented image consists of 0 or 1 which corresponds to the speed of sound of each material. 

Figure \ref{simulated}(a) includes an example of a speed of sound map. The pipeline was modelled as carbon steel ($s = 5920$m/s) and the defects and pipe wall as air ($s=343$m/s). We simulated ultrasonic data with k-Wave \cite{kwave} and used them as input for training. The respective speed of sound maps were used as targets. Each simulation took approximately 5 minutes to run on an NVIDIA Geforce GTX 970. We limited the generation to only 230 scenarios (training data: 180, test data: 50) where we randomly varied the number and location of defects in a pipeline. To increase difficulty, we added noise and under-sampled sources by a factor of two. An example of a DAS image using clean, fully-sampled data is shown in Figure \ref{simulated}(b). The defects are correctly located but there are side lobes present due to the limited spatial coverage of the linear array. Figure \ref{simulated}(c) includes the DAS image from noisy, under-sampled data. In this case, it is more challenging to localize the defects in the ultrasonic image. 

We use the noisy and under-sampled data to evaluate the three training strategies introduced in the previous section. All strategies are implemented in PyTorch, and DCNN parameters are optimized using the Adam optimization \cite{adam} with a learning rate of $10^{-3}$. The average cross entropy (lower is better) of each strategy on the test set is: $3.4\times 10^{-3}$, $6.7\times 10^{-4}$ and $1.2\times 10^{-4}$ each. A visual comparison is given in Figure \ref{simulated}(d), \ref{simulated}(e) and \ref{simulated}(f) where the segmented images obtained by each strategy are shown. These results demonstrate that the proposed end-to-end integration of the DAS operator with both data pre-processing and image post-processing steps leads to a substantial improvement of the final segmentation result.

\subsection{Real data}
\begin{figure}
\centering
\subfloat[]{ \hspace{2mm}
	\includegraphics[scale=0.184]{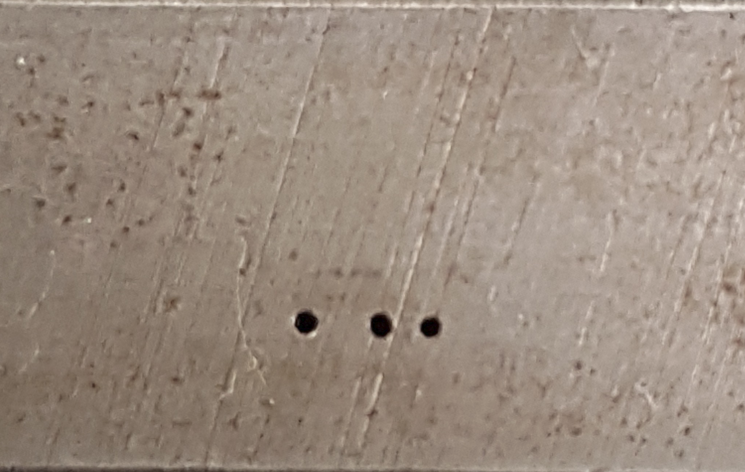}
	}\\
\subfloat[]{ 
	\includegraphics[scale=0.385]{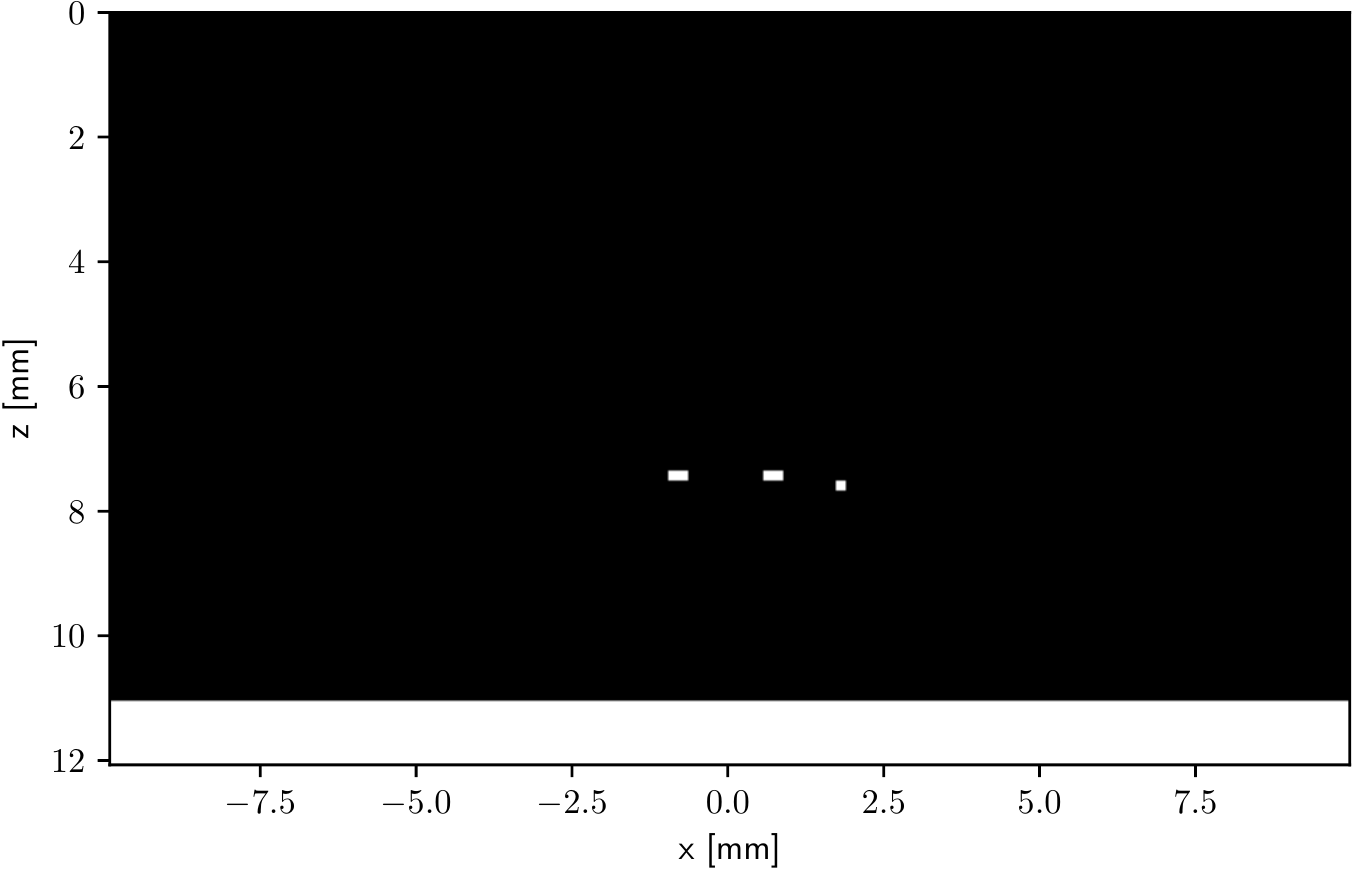}
	}\\
\caption{(a) Picture of a carbon steel block, (b) segmentation result of proposed end-to-end deep learning approach.}
\label{real}
\end{figure}

To further validate our proposed approach, we acquired real ultrasonic data using a carbon steel block with three holes. A picture can be seen in Figure \ref{real}(a). We used the data acquired (half of the sources) as input and estimated a segmented image. The spatial extent of the defects is underestimated since we did not take into account the temporal impulse response of the receivers during training data simulation. Nevertheless, we obtain an accurate localization and separation of the defects as seen in Figure \ref{real}(b). Our end-to-end deep learning approach was trained only on simulated data but it was able to transfer the learnt representations to more challenging real data.

\section{Conclusions}
Deep learning can be integrated into existing ultrasonic imaging workflows and replace traditional data pre-processing and image post-processing steps with success. Nevertheless, there are various architectures and strategies for training deep neural networks. In this work, we proposed an end-to-end approach that integrates the image formation into the network architecture. This results in a single network that maps raw data to desired imaging result. We demonstrated this concept for the DAS image formation and for segmentation as an image post-processing task. To increase difficulty, we sub-sampled the noisy data by half, which could be used to speed up the data acquisition in real-world applications. Experiments have shown that end-to-end training produces better segmented images as opposed to training for each task separately. Even though the final cost function is the same, sub-optimal results are obtained when training steps sequentially. This is because we fix the parameters of the data pre-processing network and only optimize the parameters of the image post-processing network. On the other hand, end-to-end training is more flexible since it optimizes the parameters of both networks simultaneously. It is initialized with the learnt network parameters of the sequential approach and can only improve upon those. Furthermore, training was performed only on simulated data, but the proposed approach was successful on real data illustrating the potential of deep learning to learn from physics simulations to solve real-world ultrasonic inverse problems.

\bibliographystyle{IEEEtran}

\bibliography{references}

\begin{thebibliography}{10}
\providecommand{\url}[1]{#1}
\csname url@samestyle\endcsname
\providecommand{\newblock}{\relax}
\providecommand{\bibinfo}[2]{#2}
\providecommand{\BIBentrySTDinterwordspacing}{\spaceskip=0pt\relax}
\providecommand{\BIBentryALTinterwordstretchfactor}{4}
\providecommand{\BIBentryALTinterwordspacing}{\spaceskip=\fontdimen2\font plus
\BIBentryALTinterwordstretchfactor\fontdimen3\font minus
  \fontdimen4\font\relax}
\providecommand{\BIBforeignlanguage}[2]{{%
\expandafter\ifx\csname l@#1\endcsname\relax
\typeout{** WARNING: IEEEtran.bst: No hyphenation pattern has been}%
\typeout{** loaded for the language `#1'. Using the pattern for}%
\typeout{** the default language instead.}%
\else
\language=\csname l@#1\endcsname
\fi
#2}}
\providecommand{\BIBdecl}{\relax}
\BIBdecl

\bibitem{ultrafast}
M.~{Tanter} and M.~{Fink}, ``Ultrafast imaging in biomedical ultrasound,''
  \emph{IEEE Transactions on Ultrasonics, Ferroelectrics, and Frequency
  Control}, vol.~61, no.~1, pp. 102--119, 2014.

\bibitem{dlus}
R.~J.~G. {van Sloun}, R.~{Cohen}, and Y.~C. {Eldar}, ``Deep learning in
  ultrasound imaging,'' \emph{Proceedings of the IEEE}, vol. 108, no.~1, pp.
  11--29, 2020.

\bibitem{acb}
S.~{Khan}, J.~{Huh}, and J.~C. {Ye}, ``Adaptive and compressive beamforming
  using deep learning for medical ultrasound,'' \emph{IEEE Transactions on
  Ultrasonics, Ferroelectrics, and Frequency Control}, 2020.

\bibitem{dlfab}
B.~{Luijten}, R.~{Cohen}, F.~J. {de Bruijn}, H.~A.~W. {Schmeitz}, M.~{Mischi},
  Y.~C. {Eldar}, and R.~J.~G. {van Sloun}, ``Deep learning for fast adaptive
  beamforming,'' in \emph{2019 IEEE International Conference on Acoustics,
  Speech and Signal Processing (ICASSP)}, 2019, pp. 1333--1337.

\bibitem{perdios}
D.~{Perdios}, M.~{Vonlanthen}, F.~{Martinez}, M.~{Arditi}, and J.~{Thiran},
  ``Deep learning based ultrasound image reconstruction method: A time
  coherence study,'' in \emph{2019 IEEE International Ultrasonics Symposium
  (IUS)}, 2019, pp. 448--451.

\bibitem{simson}
\BIBentryALTinterwordspacing
W.~Simson, R.~G{\"{o}}bl, M.~Paschali, M.~Kr{\"{o}}nke, K.~Scheidhauer,
  W.~Weber, and N.~Navab, ``End-to-end learning-based ultrasound
  reconstruction,'' \emph{CoRR}, vol. abs/1904.04696, 2019. [Online].
  Available: \url{http://arxiv.org/abs/1904.04696}
\BIBentrySTDinterwordspacing

\bibitem{segmNair}
A.~A. {Nair}, K.~N. {Washington}, T.~D. {Tran}, A.~{Reiter}, and M.~A.~L.
  {Bell}, ``Deep learning to obtain simultaneous image and segmentation outputs
  from a single input of raw ultrasound channel data,'' \emph{IEEE Transactions
  on Ultrasonics, Ferroelectrics, and Frequency Control}, 2020.

\bibitem{ongie2020deep}
G.~Ongie, A.~Jalal, C.~A. Metzler, R.~G. Baraniuk, A.~G. Dimakis, and
  R.~Willett, ``Deep learning techniques for inverse problems in imaging,''
  2020.

\bibitem{arridge}
S.~Arridge, P.~Maass, O.~\"{O}ktem, and C.-B. Sch\"{o}nlieb, ``Solving inverse
  problems using data-driven models,'' \emph{Acta Numerica}, vol.~28, pp.
  1--174, 2019.

\bibitem{tfm}
C.~Holmes, B.~W. Drinkwater, and P.~D. Wilcox, ``Post-processing of the full
  matrix of ultrasonic transmit receive array data for non-destructive
  evaluation,'' \emph{NDT \& E International}, vol.~38, no.~8, pp. 701 -- 711,
  2005.

\bibitem{iwex}
N.~{Portzgen}, D.~{Gisolf}, and G.~{Blacquiere}, ``Inverse wave field
  extrapolation: a different {NDI} approach to imaging defects,'' \emph{IEEE
  Transactions on Ultrasonics, Ferroelectrics, and Frequency Control}, vol.~54,
  no.~1, pp. 118--127, 2007.

\bibitem{ws}
\BIBentryALTinterwordspacing
S.~Qiao, H.~Wang, C.~Liu, W.~Shen, and A.~L. Yuille, ``Weight
  standardization,'' \emph{CoRR}, vol. abs/1903.10520, 2019. [Online].
  Available: \url{http://arxiv.org/abs/1903.10520}
\BIBentrySTDinterwordspacing

\bibitem{gn}
Y.~Wu and K.~He, ``Group normalization,'' in \emph{The European Conference on
  Computer Vision (ECCV)}, September 2018.

\bibitem{unet}
O.~Ronneberger, P.~Fischer, and T.~Brox, ``U-net: Convolutional networks for
  biomedical image segmentation,'' in \emph{Medical Image Computing and
  Computer-Assisted Intervention -- MICCAI 2015}, 2015, pp. 234--241.

\bibitem{kwave}
\BIBentryALTinterwordspacing
B.~E. Treeby and B.~T. Cox, ``{k-Wave: MATLAB toolbox for the simulation and
  reconstruction of photoacoustic wave fields},'' \emph{Journal of Biomedical
  Optics}, vol.~15, no.~2, pp. 1 -- 12, 2010. [Online]. Available:
  \url{https://doi.org/10.1117/1.3360308}
\BIBentrySTDinterwordspacing

\bibitem{adam}
D.~P. Kingma and J.~Ba, ``Adam: A method for stochastic optimization,'' in
  \emph{3rd International Conference for Learning Representations}, 2015.

\end{thebibliography}

\end{document}